\documentclass[aps,prc,twocolumn, superscriptaddress,showkeys,nofootinbib]{revtex4-1}  

\usepackage[utf8]{inputenc}
\usepackage{graphicx}
\usepackage{subcaption}

\usepackage{graphicx,color}
\usepackage{amsmath,amssymb,amsfonts}
\usepackage{hyperref}
\usepackage{xspace}	

\begin{document}
\title{
A Fixed-Target Experiment at the Muon Collider}
\medskip
\author{Henry~T.~Klest} 
\email{hklest@anl.gov}
\affiliation{Physics Division, Argonne National Laboratory, \\
Lemont, IL 60439, USA}

\begin{abstract}
The Muon Collider, recently highlighted as Recommendation 1 in the U.S. National Academies report on Elementary Particle Physics, offers a unique opportunity for fixed-target experiments with high energy and luminosity. This paper outlines some of the challenges and possibilities for fixed-target experiments to study the multi-dimensional structure of hadrons at the Muon Collider. We present a sketch of an experiment making use of the high-energy muon beam from the Muon Collider that could serve as the next-generation hadronic physics experiment after the Electron-Ion Collider.
\end{abstract}
\maketitle

\section{Introduction} \label{sec:intro}

Since the pioneering deep inelastic scattering (DIS) experiments at SLAC, the repertoire of phenomena that can be used as tools to teach us about the partonic structure of visible matter has been steadily expanded by improved accelerator facilities. Lepton scattering at high energies and with high integrated luminosity provides the best opportunity to understand the partonic structure of visible matter. Processes with large cross sections have already been measured, yet our understanding of nucleon structure in terms of partons remains largely one-dimensional. The future of hadronic physics therefore lies in measuring processes with small cross sections but unique sensitivity to otherwise inaccessible physical phenomena~\cite{Achenbach:2023pba,AbdulKhalek:2021gbh}.

\begin{figure}
    \centering
    \qquad 
    \includegraphics[width=1\linewidth]{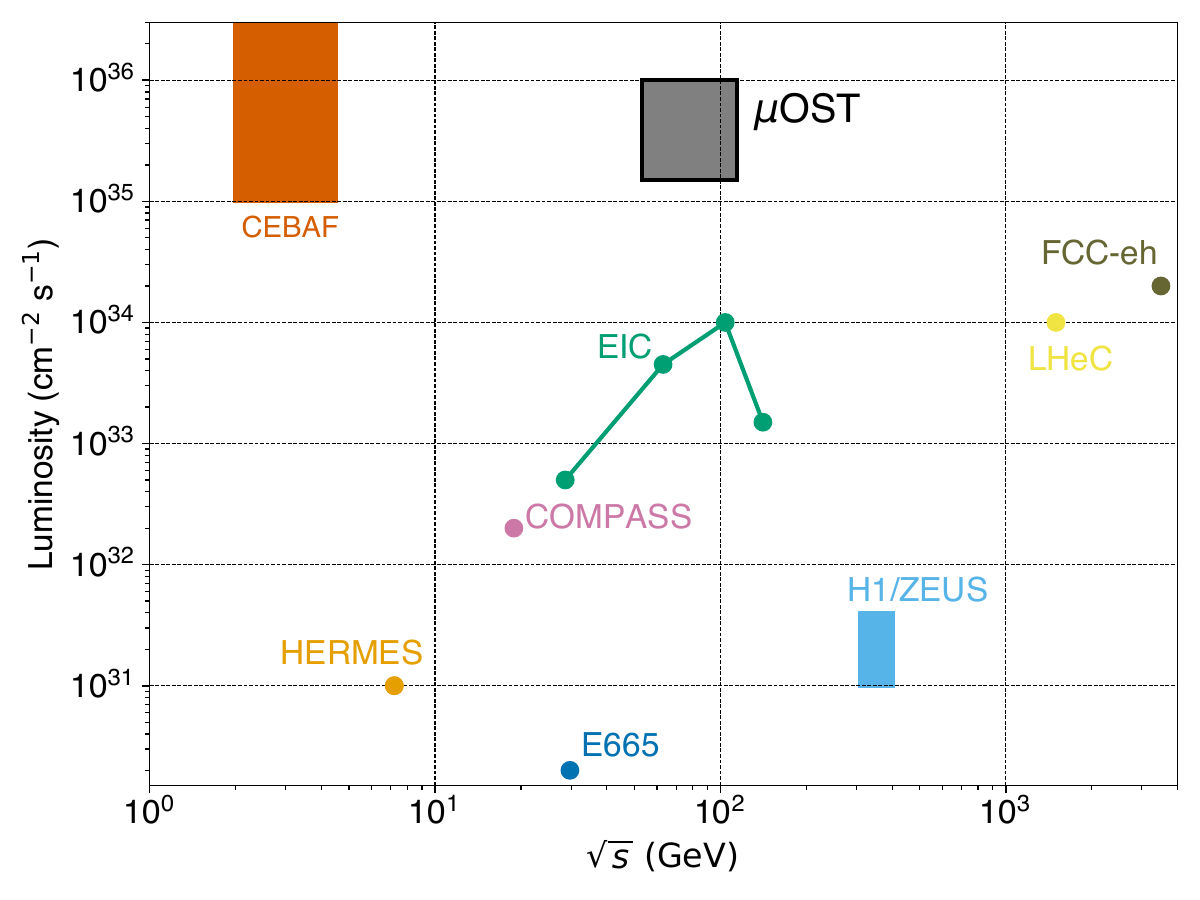}
    \caption{Comparison of luminosity and energy for lepton-scattering facilities. The $\mu$OST box corresponds to 1\% of the 1.5 to 7 TeV collider beam being dumped on a 1- to 6-meter long target.
}
    \label{fig:Lumi}
\end{figure}

Historically, energy and luminosity have been anti-correlated for lepton-scattering experiments (see Fig.~\ref{fig:Lumi}). For example, the collider experiments H1 and ZEUS at HERA operated at $\sqrt{s}=318$ GeV but were limited to around 1 fb$^{-1}$ of integrated luminosity accumulated over 15 years of running, while fixed-target experiments at much lower energies such as those at Jefferson Lab ($\sqrt{s}<5$ GeV) regularly collect a similar integrated luminosity in seconds. This is attributable mainly to the fact that even dilute gaseous targets are many orders of magnitude denser than collider beams. 

Multi-dimensional study of hadrons necessitates high luminosity to combat the curse of dimensionality. At the same time, high energy is vital to reach the regime where perturbative QCD is valid and predictive, enabling extraction of nucleon structure quantities from perturbative calculations. The instantaneous luminosity of the EIC is planned to be $10^{33}$ to $10^{34}$ cm$^{-2}$/s, which is sufficient for multi-dimensional measurements of processes with low cross sections that were inaccessible at HERA~\cite{AbdulKhalek:2021gbh}. However, this luminosity is still significantly lower than those accessible at fixed-target experiments. Additionally, the present scope of the EIC includes a single collider experiment. Since the capabilities of the EIC machine are unique, there will be no ability to independently cross check the results of that experiment unless a second detector is constructed.

With these aspects in mind, one can ask what other facilities can complement or surpass the EIC in terms of energy and luminosity. Perhaps the most similar existing facility to the EIC in terms of energy and luminosity is the M2 beamline of the CERN SPS, home to AMBER and COMPASS~\cite{Adams:2018pwt,COMPASS:2007rjf}. M2 provides a secondary beam of muons with energies up to 190 GeV ($\sqrt{s}=18.9$ GeV) and 
attained a luminosity of around $5\cdot10^{32}$ cm$^{-2}$/s for COMPASS. The highest energy fixed-target charged-lepton-scattering experiment to date was E665 at Fermilab, which operated with a muon beam of around 500 GeV~\cite{E665:1989lkq} but a luminosity below $10^{31}$ cm$^{-2}$/s. Two future proposals, the LHeC~\cite{LHeC:2020van,Bordry:2018gri} and FCC-eh~\cite{Bordry:2018gri,FCC:2018byv}, reach higher energies by adding electron rings to the LHC and FCC. Another future facility that has been discussed is a Muon-Ion Collider~\cite{Acosta:2021qpx,Acosta:2022ejc,ClineMuIC,Acosta:2025tdh,Davoudiasl:2024fiz,Apyan:2025ujv,Zhao:2025kvq,Ahluwalia:2022qsp} that could reach a $\sqrt{s}$ of 140 GeV to 6.5 TeV with a luminosity of $10^{31}$ to $10^{32}$ cm$^{-2}$/s, lower than the EIC and similar to HERA. 

The goal of this paper is to present another future alternative to complement and expand upon the physics of the EIC in the mid-21st century using the beam of the Muon Collider for a fixed-target experiment. We denote this experiment Muons On Stationary Targets, or \textit{$\mu$OST}. The Muon Collider concept has been discussed for decades~\cite{Ruggiero:1992zv,Proceedings:1993ere,Barger:1995hr,NeutrinoFactory:2005cgg,Mokhov:2011zzd,Delahaye:2013jla,Wang:2015yyh,MuonCollider:2022nsa}, culminating in positive endorsements for Muon Collider R\&D by the most recent iterations of the European Particle Physics Strategy~\cite{CERN-ESU-015-2020} group and U.S. Particle Physics Project Prioritization Panel~\cite{P5:2023wyd}. The 2025 U.S. National Academies report on elementary particle physics~\cite{NAS2025Higgs} listed Muon Collider R\&D as Recommendation 1 for U.S. particle physics in the coming decades. As the recently formed International Muon Collider Collaboration~\cite{InternationalMuonCollider:2025sys} undertakes R\&D and design studies, it is timely to consider the inherent potentiality of the Muon Collider for hadronic physics\footnote{The idea of using the Muon Collider beam as a source of \textit{neutrinos} for fixed-target experiments has been discussed widely; a specific example is the nuSTORM concept~\cite{Adey:2015iha}. High-intensity beams of TeV-scale neutrinos are also inherently powerful for hadronic structure measurements~\cite{NuTeV:2001whx,Bai:2020ukz,Anchordoqui:2021ghd,Feng:2022inv,Xie:2023suk,Dulat:2015mca,Candido:2023utz,deGouvea:2025zfq,Ariga:2025qup,InternationalMuonCollider:2025sys,Francener:2025pnr}.}.

\section{Energy and Luminosity}
Various center-of-mass energies have been explored as possible operation points for Muon Colliders, ranging from the Higgs pole at 62.5 GeV per beam to 7 TeV per beam. For simplicity, we restrict ourselves to considering the designs discussed in Ref.~\cite{Accettura:2023ked}. The 7 TeV beam provides the highest energy and lowest beam-induced background for a fixed-target experiment; however, it makes the design of the detector more challenging due to the highly boosted center-of-mass, as will be described in Sec.~\ref{detector}. The physics reach can be extended by performing measurements also at lower beam energies, where the final state is less forward-boosted.

When delivered on a nucleon target at rest, the 7 TeV beam produces a center-of-mass energy of $\sqrt{s} = \sqrt{2\,m_p\,E_{\rm beam}} = 115\ \mathrm{GeV}.$ This energy is very similar to that of the EIC, where the highest energy configuration, an 18 GeV electron on a 275 GeV proton, provides a $\sqrt{s}=141$ GeV. The highest luminosity configuration of the EIC ($10^{34}$ cm$^{-2}$/s) will be achieved at $\sqrt{s}=104$ GeV with a 10 GeV electron beam incident on a 275 GeV proton beam.

Projections for the beam conditions of the Muon Collider were provided in Ref.~\cite{Accettura:2023ked}, and we will use these numbers to make some rough estimates of what could be achieved in terms of luminosity in a fixed-target experiment. For the 7 TeV beam, the number of muons per bunch is $1.8\cdot10^{12}$ and the repetition rate is 5 Hz, corresponding to an integrated beam current of around 1.4 $\mu A$. This is about two orders of magnitude smaller than the beam current available at Jefferson Lab, but for muon scattering experiments, the target can be made very long without significantly impacting the muons as they enter and leave the target. If this beam were delivered directly onto a 1-meter-long liquid hydrogen target, it would produce approximately 3.3 ab$^{-1}$ of integrated luminosity per day, resulting in an average instantaneous luminosity of 4$\cdot 10^{37}$ cm$^{-2}$/s.


While the aforementioned luminosity scenario appears attractive at first, the high bunch charge and low repetition rate will lead to enormous pileup if the structure of the beam reaching the target is the same as it is in the collider. One bunch of $1.8\cdot10^{12}$ muons at 7 TeV passing through a 1 m long liquid hydrogen target could result in tens of thousands of simultaneous scattering events reaching the detector system. Realistically, the beam will only be dumped once a significant fraction of the muons have decayed. However, even if only 1\% of the muons sent into the collider are put on target, the potential physics case is still significant, as will be shown in Sec.~\ref{sec:physics}. 

We assume this ``1\% scenario'' for several reasons. The extraction of 1\% of the collider beam is unlikely to affect the statistical precision attainable by the collider experiments. With 1\% of the nominal collider beam current and an appropriately chosen detector acceptance, the level of pileup is similar to the HL-LHC. With the assumption of additional R\&D advancements in detector and data processing technologies driven by HL-LHC, FCC, and the Muon Collider detectors, this seems like a technically feasible operating point. The pileup issue can be further ameliorated by extracting muons for the fixed-target experiment every turn of the collider instead of dumping the beam all at once, but this adds complexity to the collider ring design. Assuming 20 weeks per year of physics running, the 1\% beam dump on a 1-meter target scenario would result in an average instantaneous luminosity of approximately 1.5$\cdot 10^{35}$ cm$^{-2}$/s, around an order of magnitude higher than the maximum luminosity of the EIC. Operation at beam energies down to 1.5 TeV should provide similar luminosities for a fixed-target experiment, due to the similar repetition rate and number of muons per bunch~\footnote{In the collider, the increase in luminosity with energy is driven by the smaller beam size and number of useful rotations, neither of which significantly affect the fixed-target luminosity.}. Furthermore, depending on the dump design, the $\mu^+$ and $\mu^-$ beams could be utilized independently by two experiments located at the $\mu^+$ and $\mu^-$ beam dumps, respectively. A longer target, such as the 6-meter-long target of EMC~\cite{EuropeanMuon:1980nje}\footnote{A more extreme example is the 40-meter target of BCDMS~\cite{Bollini:1982ac}.}, could potentially increase the luminosity to almost 1$\cdot 10^{36}$ cm$^{-2}$/s, but at the cost of higher pileup. 

The authors of Refs.~\cite{Cesarotti:2022ttv,Cesarotti:2023sje} demonstrated the impressive reach of a beam dump experiment for studying long-lived BSM particles. Since the muons are highly penetrating, such a beam dump experiment could be operated in parallel with the nuclear physics experiment described here. With the assumed 1\% scenario, the long-lived particle search experiment would reach $10^{18}$ muons on target in the course of a couple years of running.

\section{Detector, Target, and Environment}
\label{detector}
\begin{figure*}[t]
    \centering
    \qquad 
    \includegraphics[width=1\linewidth]{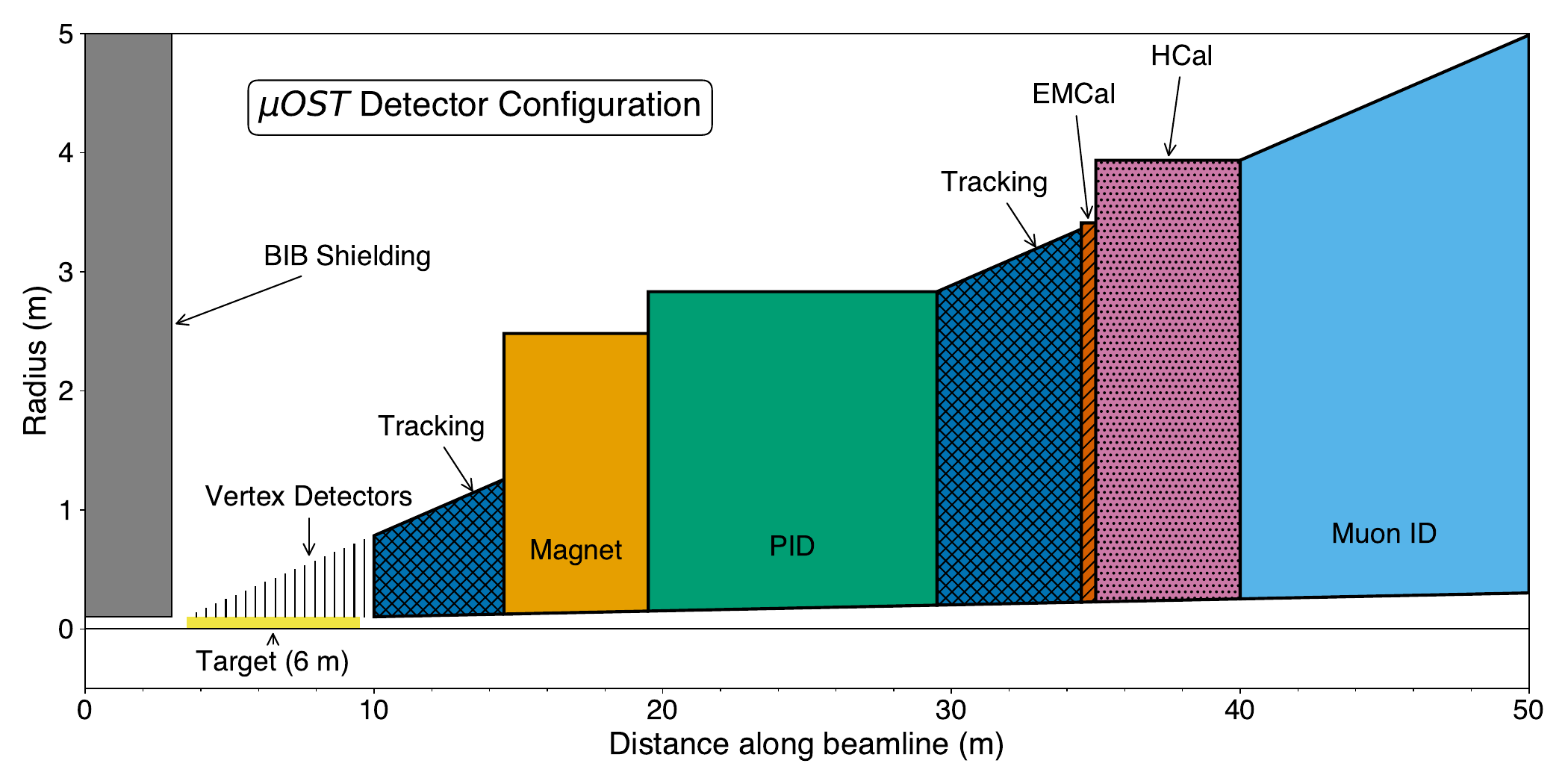}
    \caption{Example $\mu$OST detector configuration with acceptance $5 < \theta < 105$ mrad for a 6 meter long target.}
    \label{fig:Det}
\end{figure*}

Several aspects of the unique muon beam environment have significant impact on the design of the detector systems. One can imagine the detector taking several forms. A large-aperture forward detector such as LHCb or COMPASS is an obvious baseline. We assume this experiment could not be built in the collider ring à la LHCb or HERMES due to the immense beam-induced background and the space required to precisely momentum analyze the high-energy scattered muons, which scatter at small angles even at high momentum transfer, $Q^2$. Therefore, we assume that the detector resides in a long hall with a dedicated beamline in the manner of COMPASS~\cite{COMPASS:2007rjf} or SeaQuest~\cite{SeaQuest:2017kjt}. A rough sketch of such a detector with particle identification (PID) via Cherenkov is shown in Fig.~\ref{fig:Det}.

Perhaps the most obvious challenge is that at 7 TeV the lab-frame angles of produced particles are extremely small even at high $Q^2$. Therefore it seems necessary to utilize a dipole magnet that can appreciably bend particles at small scattering angles. The momentum resolution of a detector with a dipole analyzing magnet can be estimated as $\frac{dp}{p}=\sigma_x\cdot\frac{p}{h\int B\cdot dL}$, where $\sigma_x$ is the spatial resolution of the detector, $h$ is the distance from the magnet at which the measurement is performed, $B$ is the magnetic field strength, and $L$ is the distance traveled in the magnetic field. To have acceptable resolution on the DIS variables within a reasonable range in inelasticity $y$, the momentum and angle of the scattered muon should be measured precisely even at TeV energies. LHCb achieves momentum resolution of around 1\% for 200 GeV particles with a $\int B\cdot dL$ of 4 Tm; to achieve a similar momentum resolution for 2 TeV particles with the same detector would require a $\int B\cdot dL$ of 40 Tm. Perhaps a more practical solution is to increase the lever arm, $h$, compared to LHCb, which is around 20 m long. For a longer spectrometer, it could be feasible to have a dipole magnet of smaller angular aperture than LHCb with $\int B\cdot dL$ of 20 Tm and a factor of two larger lever arm. The dipole could be made superconducting to reach higher field, at the cost of increased engineering complexity. This momentum resolution coupled with a scattering angle resolution of 2\%, similar to what has been achieved in COMPASS~\cite{Adams:2018pwt}, would enable measurements of $Q^2$ at the 4\% level in the region $Q^2 > 100$ GeV$^2$ for the 7 TeV beam.


The beam-induced background (BIB) plays a dominant role in the design of the detector for Muon Collider experiments. In the collider, this background arises from secondary particles produced by decay electrons interacting with the beamline and surrounding materials. For 7 TeV muons at the nominal collider beam current of $2\cdot10^{12}$ per bunch, there will be approximately $6\cdot10^4$ decays per meter~\cite{Accettura:2023ked}, rising to $2\cdot10^5$ at 1.5 TeV. Fixed-target experimental halls are typically on the order of 20-50 m long. This means that the detector can be shielded from BIB produced upstream of the target with several meters of shielding. Depending on the distance to the experimental hall from the collider ring, it could be preferable to have a final bending magnet immediately upstream of the experiment that bends the primary beam into the experimental hall and dumps the decay electrons, which carry approximately one-third of the beam momentum on average. The BIB that may be challenging to completely mitigate is the decay electrons produced inside and after the target. These electrons can be deflected out of the beamline and into the detector acceptance by the dipole magnet. For this reason, it is likely necessary to magnetically shield the primary beamline from the detector dipole as was done in HERMES~\cite{HERMES:1998mat}, at the cost of experimental acceptance.




Compared to existing targets, the requirements on the target for $\mu$OST are somewhat unique. In the case of the slowly extracted muon beam, standard nuclear or hydrogen targets should be sufficient. However, for the beam dump option, the target will need to withstand on the order of $2\cdot10^{10}$ incident muons arriving nearly simultaneously but at a low repetition rate. This challenge may be alleviated somewhat by the fact that the beam is allowed to be reasonably wide upon arrival. Compared to electron beam experiments where the electrons are easily disrupted by target material as they leave the target cell, the scattered muons can traverse more target material without their trajectories being appreciably altered. Therefore, a target with cross sectional area of $\mathcal{O}~\text{cm}^2$ accepting a wide beam is a straightforward option to reduce hotspots in the target. With sufficient vertex resolution and ability to handle pileup, multiple targets could be operated in parallel without significant disruption to the beam.

Vertexing, fast timing, and advanced data acquisition techniques such as online reconstruction for triggering will be crucial to mitigate pileup and BIB, particularly for longer and wider target options. Fast-timing detectors will be able to reject hits from out-of-time BIB and reduce the occupancy; significant R\&D and experience in this direction can be inherited from future collider experiments~\cite{MuonCollider:2022glg,MAIA:2025hzm,InternationalMuonCollider:2025sys}. For heavy-flavor measurements, the produced bottom or charm mesons will likely decay inside the target, meaning the vertex detector should ideally be able to distinguish displaced vertices inside the target. The LHCb collaboration has demonstrated software-based triggering capable of implementing complex analysis cuts at the trigger-level~\cite{LHCb:2018zdd}. Since $l+p$ scattering events are kinematically over-constrained when the hadronic final state is measured, this kind of streaming data acquisition pipeline will be an extremely powerful tool to maximize the amount of useful data collected in a harsh environment.

One optional configuration for the detector system is to employ a technique often used in hadron beam Drell-Yan experiments, such as SeaQuest~\cite{SeaQuest:2017kjt}, whereby thick absorbers filter out all final-state particles except for muons. This could be a more cost-effective day-one configuration of the detector with a full LHCb-style detector being implemented later on. In the detector design with an absorber, scattered muons and muons from hadron decays would have significantly lower backgrounds. Triple-muon events could be used to study highly sought-after processes such as double-DVCS~\cite{Guidal:2002kt} and electroproduction of vector mesons, e.g. $J/\psi$ and $\Upsilon$, for imaging the spatial distributions of partons inside of nucleons and nuclei. \begin{figure*}[t] 
  \centering
  \begin{subfigure}[t]{0.48\textwidth}
    \centering
    \includegraphics[%
        width=1\linewidth,
        trim={0.4cm 0.5cm 1cm 1cm},
        clip
    ]{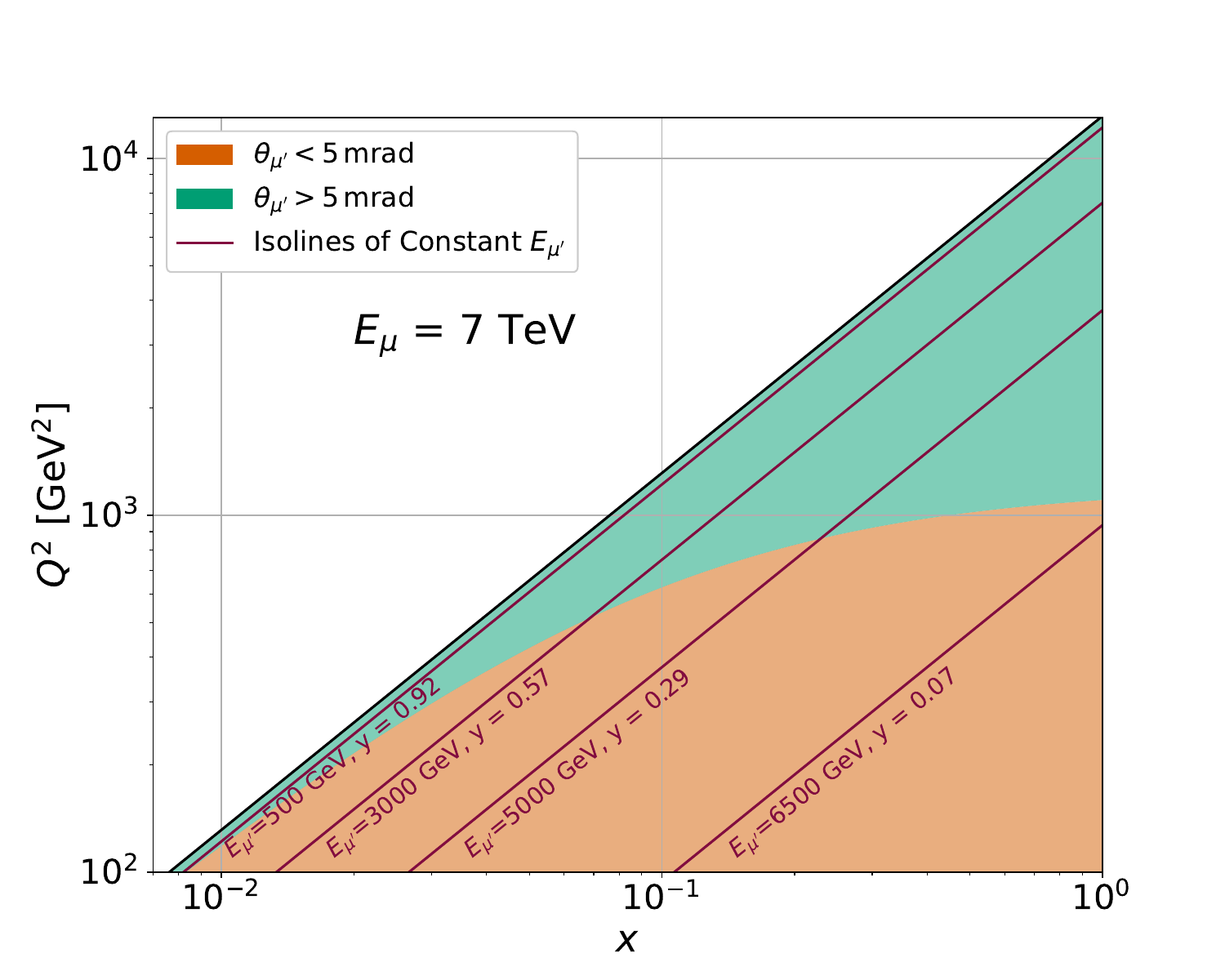}
    \label{fig:DISPlane7TeV}
  \end{subfigure}
  \hfill
  \begin{subfigure}[t]{0.48\textwidth}
    \centering
    \includegraphics[%
        width=1\linewidth,
        trim={0.4cm 0.5cm 1cm 1cm},
        clip
    ]{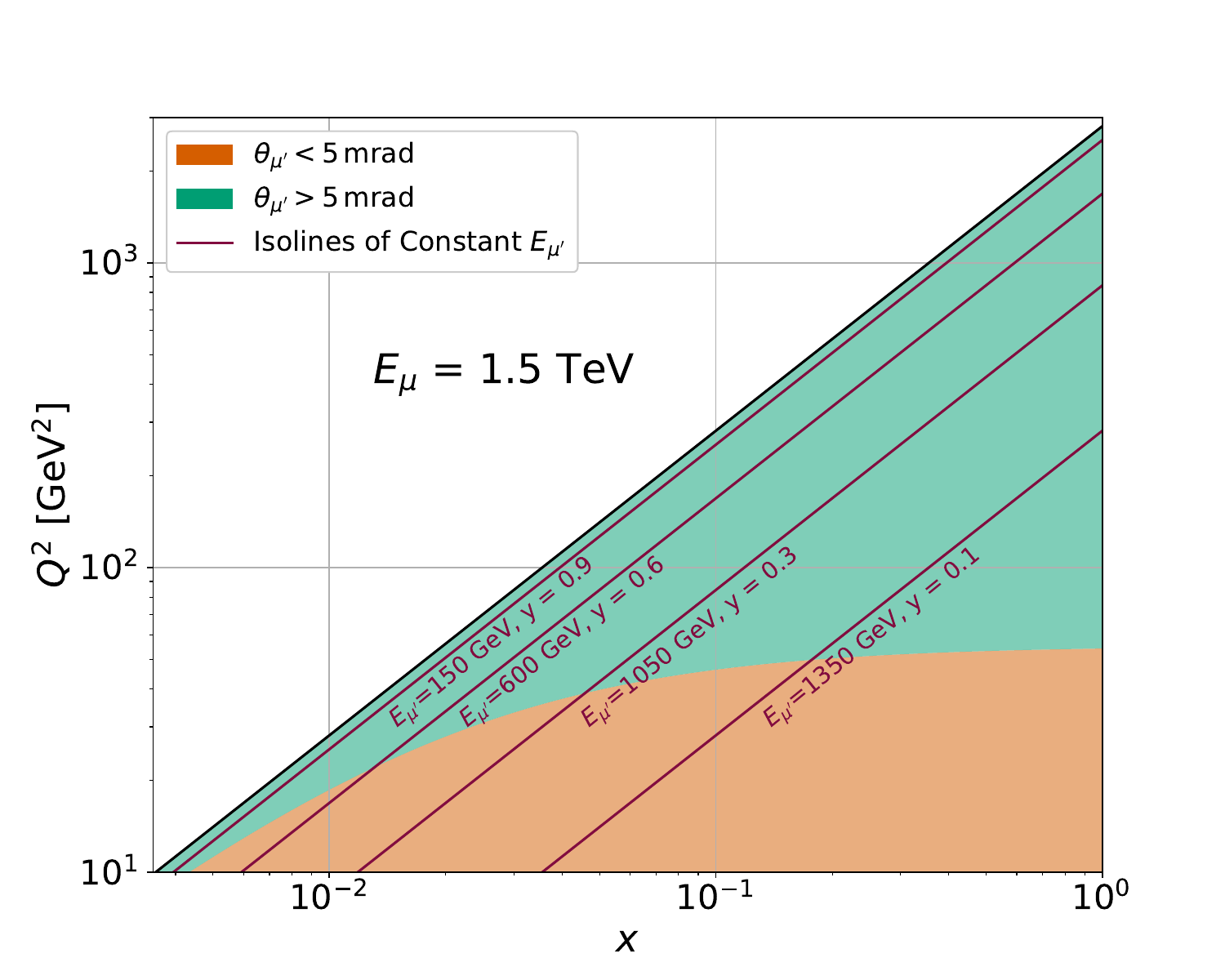}
    \label{fig:DISPlane1TeV}
  \end{subfigure}
  \caption{DIS kinematic plane for 7 TeV and 1.5 TeV beam energies. The kinematic region accessible to a detector with acceptance $\theta_{\mu'}>5$ mrad is highlighted in green. Note the different $y$-axis scales.}
  \label{fig:DISPlaneCompare}
\end{figure*}

\section{Physics Case} \label{sec:physics}

The $\mu$OST facility offers a rich physics program that spans a wide range of QCD and BSM phenomena. In contrast to the EIC, which will provide access to the low-$x$ regime of nucleon and nuclear structure, $\mu$OST can deliver high-statistics measurements at higher $Q^2$ and moderate-to-large $x$ as shown in Figs.~\ref{fig:DISPlaneCompare}. $\mu$OST provides unique opportunities to study the valence structure of the nucleon, rigorously utilize perturbative QCD in a regime where non-perturbative corrections are suppressed, and explore new physics with unprecedented sensitivity. In the following subsections we touch briefly upon several potential topics that could be studied at the $\mu$OST facility.

\subsection{Precision QCD and Parton Distributions}

With its high instantaneous luminosity and high beam energy, $\mu$OST opens the door to a unique regime for measurements of inclusive deep inelastic scattering (DIS). $\mu$OST will provide ample data for extraction of the valence region parton distribution functions (PDFs) with reduced theoretical uncertainties arising from higher-twist effects and target-mass corrections. Measurements at $Q^2$ values up to several thousand GeV$^2$ will provide critical input to global PDF fits, especially in the high-$x$ region~\cite{Burkert:2022hjz}. With a high-power polarized target, the helicity PDFs can be measured with high precision. The high $Q^2$ reach of $\mu$OST minimizes non-perturbative contamination and facilitates clean tests of spin sum rules in QCD. 

Furthermore, DIS structure functions have historically provided some of the most powerful constraints on the value of $\alpha_s$~\cite{dEnterria:2022hzv}. Since polarized and nuclear targets can also be leveraged to exploit various QCD sum rules and reduce the typically dominant theoretical uncertainties~\cite{Kutz:2024eaq}, an extraction of $\alpha_s$ at the level of 1\% or better from $\mu$OST alone seems feasible, to be compared with the global uncertainty of 0.8\% from all world measurements combined~\cite{dEnterria:2022hzv}.

Measurements of parity-violating asymmetries in DIS, which increase in magnitude with $Q^2$, provide sensitivity not only to the strangeness content of the nucleon and the $u/d$ ratio but also to potential BSM contributions. The high-statistics of $\mu$OST would enable precision asymmetry measurements at $x\approx10^{-1}$ and $Q^2\approx1000$ GeV$^2$, where electroweak interference can be used as a tool to gain new insights into valence-region QCD dynamics.

\subsection{Semi-Inclusive and Jet Physics}

With $\mu$OST, semi-inclusive DIS (SIDIS) can be measured in the regime where transverse momentum distributions of partons can be rigorously extracted without substantial theoretical uncertainty or bias. At present the vast majority of data in the valence region cannot be reliably used to study TMDs. Measuring SIDIS at higher $Q^2$ is critical for understanding the three-dimensional structure of nucleons beyond collinear factorization, and the necessary set of measurements are extremely well-suited to the $\mu$OST setup. 

For the 7 TeV beam energy at $x>0.1$ and $Q^2\gtrsim600$ GeV$^2$, the Born-level struck parton is produced at large enough angles to enter the detector. With approximately 50 million of these events per year, the ongoing program of jet physics at HERA and the EIC could be extended to the valence region. With precise vertexing, heavy flavor jets could provide direct evidence for intrinsic charm or bottom in the valence structure of the proton.

\subsection{Exclusive Measurements}

Much of the exciting physics at the EIC is driven by multi-dimensional measurements of exclusive processes, which enables "imaging" of hadrons in position space. Since exclusive cross sections usually fall strongly with $Q^2$, $\mu$OST with a 7 TeV beam is suboptimal. However, at a beam energy of 1.5 TeV, the scattered muon can be measured down to $Q^2$ values of $10$ to $20$ GeV$^2$ at an energy similar to the EIC at $10$x$100$ GeV but with almost two orders of magnitude higher luminosity. Measurements of Deeply Virtual Compton Scattering and Deeply Virtual Meson Production at perturbative values of $Q^2$ and $x>10^{-2}$ can complement EIC results at lower-$x$ on the spatial and gravitational structure of nucleons and nuclei. 
\flushbottom

Existing data does not permit the extraction of GPDs, as the measured experimental observables contain combinations of GPDs that cannot yet be uniquely disentangled from one another~\cite{Bertone:2021yyz,Moffat:2023svr}. The field is in need of complementary observables that enable this disentanglement, and beam charge asymmetry measurements are one of the pillars of the CEBAF positron upgrade for this reason~\cite{CLAS:2023gja,Zhao:2021zsm,Accardi:2020swt}. If the Muon Collider complex is constructed such that both muons and antimuons can be delivered to the same experiment, multi-dimensional beam charge asymmetries of exclusive processes such as double-DVCS can be measured at perturbative scales to help solve the problem of extracting GPDs~\cite{Moffat:2023svr}. $\mu$OST is particularly well-suited for double-DVCS by virtue of the high luminosity and ability to precisely measure either muons or electrons from the decay of the virtual photon.


\subsection{Nuclear Partonic Structure and QCD Dynamics}

While much of the focus thus far has been on proton structure, $\mu$OST can also provide new insights into the partonic structure of nuclei. Depending on the capability of the detector to handle pileup in denser nuclear targets, the attainable luminosity can be significantly higher than the EIC. High-statistics DIS on a variety of nuclear targets will enable detailed studies of nuclear parton distribution functions (nPDFs). In particular, $\mu$OST can probe the transition from the nucleon to the nucleus by comparing observables across targets, thereby shedding light on shadowing, anti-shadowing, and EMC effects at high $Q^2$. By comparing SIDIS and jet measurements in nuclei versus free nucleons, $\mu$OST can investigate phenomena such as color transparency and modifications of hadronization in a nuclear medium.

\subsection{BSM Searches}

A further exciting opportunity offered by $\mu$OST is the ability to perform high-precision tests of the Standard Model in an unexplored kinematic regime. With billions of DIS events at $Q^2>1000$ GeV$^2$, $\mu$OST can perform stringent tests of lepton universality, comparing muon-induced scattering to electron-induced processes. The high $Q^2$ regime is especially sensitive to contact interactions that could arise from new physics beyond the Standard Model. In the language of Standard Model Effective Field Theory (SMEFT), precise measurements of differential cross sections and asymmetries can constrain higher-dimensional operators sensitive to new physics~\cite{Boughezal:2020uwq,Bissolotti:2023vdw,Boughezal:2022pmb,Cirigliano:2021img}. This sensitivity is complementary to that provided by collider experiments, offering independent tests of BSM scenarios. 

Finally, $\mu$OST’s kinematics provide an ideal environment to study electroweak interference effects in DIS~\cite{Proceedings:2012ulb,H1:2015ubc,H1:2018mkk}, which can yield precision determinations of the weak mixing angle at momentum transfers far larger than existing measurements in lepton scattering.

\section{Conclusion}
The high-energy, high-luminosity $\mu$OST concept provides a new avenue for studying the partonic structure of nucleons and nuclei, complementary to the physics program of the EIC. If the challenges related to pile-up and beam-induced background can be overcome, the $\mu$OST facility offers the potential for high statistics measurements at high $Q^2$ and $x$ to multi-dimensionally map the valence region at unambiguously perturbative scales. Assuming the Muon Collider comes to fruition, $\mu$OST could provide a natural future direction for the international hadronic physics community after the EIC.
\section*{Acknowledgments}
We thank Don Geesaman, Sylvester Joosten, and Maria Żurek for helpful discussions and suggestions. We also thank Lindsay DeWitt for editing. This work was supported by the U.S. Department of Energy under Contract No. DE-AC02-06CH11357.

\bibliography{ref.bib} 
\end{document}